\documentclass[twocolumn, 10pt]{IEEEtran}
\IEEEoverridecommandlockouts
\usepackage{multirow}
\usepackage{tabu}
\usepackage{url}
\usepackage{array}
\usepackage{hyperref}
\newcolumntype{P}[1]{>{\centering\arraybackslash}p{#1}}
\newcolumntype{M}[1]{>{\centering\arraybackslash}m{#1}}
\newcolumntype{B}[1]{>{\centering\arraybackslash}b{#1}}
\usepackage{amsmath,amssymb,amsfonts}
\usepackage{algorithmic}
\usepackage{graphicx}
\usepackage{textcomp}
\usepackage{gensymb}
\usepackage{tikz}
\usepackage{soul}
\usepackage{xcolor}
\usepackage{cite}
\usepackage[separate-uncertainty = true,multi-part-units = repeat, load-configurations = abbreviations]{siunitx}
\usepackage{siunitx}

\begin{document}

\title{Cellular-Connected Wireless Virtual Reality: Requirements, Challenges, and  Solutions
}
\author{
\IEEEauthorblockN{Fenghe Hu, \emph{Student Member, IEEE}, Yansha Deng, \emph{Member, IEEE}, Walid Saad, \emph{Fellow, IEEE}, Mehdi Bennis, \emph{Senior Member, IEEE}, and A. Hamid Aghvami, \emph{Fellow, IEEE}}

\thanks{F. Hu, Y. Deng, and A. H. Aghvami are with the Department of Informatics, King's College London, London WC2R 2LS, UK (E-mail: \href{mailto:Fenghe.hu, yansha.deng, hamid.aghvami@kcl.ac.uk}{fenghe.hu, yansha.deng, hamid.aghvami@kcl.ac.uk}).(Corresponding author: Yansha Deng.)}
\thanks{M. Bennis is with Centre for Wireless Communications, University of Oulu, Finland, (Email: \href{mailto:bennis@ee.oulu.fi}{bennis@ee.oulu.fi}).}
\thanks{W. Saad is with Wireless@VT, Bradley Department of Electrical and Computer Engineering, Virginia Tech, Blacksburg, VA, USA, (Email: \href{mailto:walids@vt.edu}{walids@vt.edu}). This research was supported by the U.S. National Science Foundation under Grant CNS-1836802.}
}

\maketitle

\begin{abstract}
Cellular-connected wireless connectivity provides new opportunities for virtual reality (VR) to offer seamless user experience from anywhere at anytime. To realize this vision, the quality-of-service (QoS) for wireless VR needs to be carefully defined to reflect human perception requirements.  
In this paper, we first identify the
primary drivers of VR systems, in terms of applications and
use cases. We then map the human perception requirements to 
 corresponding QoS requirements for four phases of VR technology development. To shed light on how to provide short/long-range mobility for VR services, we further list four main use cases for cellular-connected wireless VR and identify their unique research challenges along with their corresponding enabling technologies and solutions in 5G systems and beyond. 
Last but not least, we present a case study to demonstrate the effectiveness of our proposed solution and the unique
QoS performance requirements of VR transmission compared with that of traditional video service in cellular networks.

\end{abstract}

\section{Introduction}
Virtual reality (VR) will revolutionize the human interactions between 
humans and their world by connecting people across global communities within highly interactive virtual components or worlds that transcend geographical boundaries. This vision has inspired the commercial release of various hardware devices, open-access standards, and APIs by various global enterprises, including ARKit from Apple, HoloLens from Microsoft, Oculus series from Facebook, and Vive series from HTC. Indeed, it is anticipated that the global VR market size will reach \textcolor{black}{\$80.7 billion by 2024}, stemming from a diverse set of VR services and devices that are expected to stimulate our human senses with realistic feedback. 


\textcolor{black}{Virtual reality (VR) is a transformative service designed to build a synthetic virtual environment}
to mimic the real world and, subsequently, immerse participants in highly realistic virtual worlds. VR will in fact provide innovative ways for users to interact with their world for various purposes such as personal entertainment and social interactions. Indeed, VR will have a very rich application domain ranging from automotive video streaming, social sharing at crowded venues, and 6 degree-of-freedom content streaming, to remote control/tactile Internet \cite{Quacom}. However, the poor user experience provided by traditional computer-supported VR headsets or all-in-one headsets (e.g., Oculus Go) limits the imagination and virtual world potential for what is possible with VR (as shown in Table. \ref{headset-parameter}).

One of the main barriers facing the mass marketization of these VR applications is the restricted mobility of wired VR devices and the lack of real-time high-quality content support of the all-in-one VR devices. To overcome these limitations, one can support VR devices with wireless cellular connectivity that can provide ubiquitous user experiences
from anytime and anywhere and can potentially unleash a plethora of novel VR applications. While \emph{cellular-connected VR} has a very promising potential, it is imperative to address many unique challenges that are not faced in wired VR systems and conventional wireless video streaming systems. Such challenges include providing seamless service over unstable wireless channels, solving handover issues when VR users are mobile, managing the asymmetric and coupled traffic in the uplink and downlink, and providing real-time VR content service \cite{Tan2018,Cuervo2018,HuaweiiLab2017}. With the growing interest in wireless VR \cite{Bastug2017a,Elbamby2018a,Saad2019}, there is a need for a concrete vision on how one can integrate wireless VR services over cellular networks. Although some recent articles such as \cite{Bastug2017a,Elbamby2018a,Saad2019} have attempted to discuss these challenges, they failed to provide quantitative quality-of-service (QoS) requirements from human perceptions and cellular-connected VR use cases perspectives. \textcolor{black}{In \cite{Saad2019}, we have provided a new definition, called quality-of-physical-experience, which jointly considers the physical fact of human itself and network QoS. Meanwhile, in \cite{Chen2018}, we have introduced the concept of breaks-in-presence to capture the QoS of wireless VR.}

The main contribution of this paper is a comprehensive study of the unique challenges and potential solutions for cellular-connected wireless VR networks. From the perspective of satisfying human perceptions, we first define human perceptions of VR services which are then mapped to unique QoS requirements for immersive VR in Section II. In particular, we categorize the development of VR QoS into four phases with their corresponding QoS requirements as well as VR device parameters. Then, we introduce the four use cases of VR defined by Qualcomm and their specific VR QoS requirements for cellular-connected VR and we identify the challenges in meeting these QoS requirements, while outlining promising solutions in Section III. We then conclude with a practical simulation to capture the challenge of supporting cellular-connected VR with the current state-of-art cellular network in Section IV.

\begin{table}[ht]
\renewcommand\arraystretch{1.1}
\centering
\label{headset-parameter}
\caption{Current MR/VR Headsets Technical Parameters}
\begin{tabular}{||c|M{1.3cm}|c|M{1.5cm}|M{1.3cm}||}
\hline
\textbf{Devices} & \textbf{FoV (diagonal)} & \textbf{PPD} & \textbf{Resolution (single-eye)} & \textbf{Refresh Rate} \\ \hline
Hololens 2  & 52\degree & 47 & / & / \\ \hline
Vive & 110\degree & 11 & 1080$\times$1200 & \SI{90}{\hertz}  \\ \hline
Pimax 8K  & 150\degree & 14 & 3840$\times$2160 & \SI{80}{\hertz} \\ \hline
Odyssey & 110\degree & 14 & 1440$\times$1600 & / \\ \hline
Oculus Rift & 110\degree & 11 & 1080$\times$1200 & \SI{90}{\hertz} \\ \hline
Playstation VR & 100\degree & 11 & 960$\times$1080 & \SI{120}{\hertz} \\ \hline
Vive Pro Eye & 110\degree & 14 & 1440$\times$1600 & \SI{90}{\hertz} \\ \hline
\end{tabular}
\label{devices}
\end{table}


\section{VR Service Requirement}
In order to define the VR QoS requirements of cellular-connected VR devices, 
the service requirements for a fully immersive virtual experience will be first discussed based on the human perception requirements, given that the performance of VR services is highly determined by the users' quality-of-physical-experience (QoPE) \cite{Saad2019}. 
Then, the QoS requirements for four VR phases pertaining to the evolution of VR devices will be detailed.


\subsection{Human Perception}
In VR applications, human perceptions can be evaluated based on resolution, field-of-view (FoV), refresh rate, and VR interaction latency. These factors will impact the QoS requirement needed for having an immersive VR experience in different applications and for various users. Human perception can vary between individuals depending on a variety of human factors that \textcolor{black}{include age, health, and occupation, among others.} In the following, we provide the average human perception values for a typical, healthy young person between 12 and 18 years old.

\subsubsection{Resolution}
The minimum VR resolution can be determined by the maximum acuity of human eyes.
For an environment with a fine black line on an illuminated white background, the maximum acuity corresponds to a minimum resolvable feature or gap of 0.3\texttildelow 1 arc-minutes, a minimum detectable object of 0.5 arc-second, a minimum visible position shift of 5 arc-second, and a minimum identifiable object of 40\texttildelow 60 arc-second. In this case, an image resolution with more than 720 pixel-per-degree (PPD) is required to fully satisfy the human eye's requirement. For a non-ideal environment with relatively low disparity scenes, a VR resolution with more than 60 PPD is needed to fully satisfy the human eye's requirement \cite{Cuervo2018}. However, as shown in Table \ref{headset-parameter}, existing commercial VR headsets can only achieve a maximum of 14 PPD  resolution, leading to the so-called ``screen door effect", in which the fine lines separating pixels become visible.

\subsubsection{Field-of-View}
The area of vision in the human eye at a specific time is known as the field-of-view (FoV) shown in \cite[Fig. 4.1]{3GPP201807}, which specifies a  210$\degree\times$150$\degree$  (i.e., diagonal 200$\degree$) FoV requirement. The FoV can be further divided into several regions with different resolution requirements. The central vision zone is the most sensitive zone with 60$\degree$ around the center. Meanwhile, peripheral vision zone is a less sensitive zone with 30$\degree$ around the central vision, and the monocular vision zone is the rest of the vision. Usually, the central vision zone has the highest resolution requirement, and the monocular vision has the lowest resolution requirement \cite[Fig. 2]{Bastug2017a}. Note that existing commercial VR headsets can only achieve a maximum 150$\degree$ of FoV as shown in Table \ref{headset-parameter}.  

\subsubsection{Refresh Rate}
The refresh rate is the number of frames per second shown by the display. A higher refresh rate leads to more continuity in the motion. A low refresh rate can cause motion blur. Depending on either the visual acuity or the motion continuity, lower bounds of the refresh rate can be obtained.  To ensure the motion continuity from human perceptions, the minimum refresh rate for a computer-rendered video is \SI{120}{\hertz} \cite{Bastug2017a}. To ensure the movement is presented pixel by pixel continuously while moving at a speed of 30 degree/second speed on a 60 PPD resolution monitor, the minimum refresh rate can be very stringent and as strict as \SI{1800}{\hertz} \cite{Cuervo2018}. Note that existing commercial VR headsets can only achieve a maximum \SI{120}{\hertz}  refresh rate as shown in Table \ref{headset-parameter}.  

\begin{table*}[htbp!]
\renewcommand\arraystretch{1.5}
\centering
\caption{QoS Requirements for VR Phases}
\label{requiremnt-different-phase}
\begin{tabular}{||m{1cm}|M{3cm}|M{2cm}|M{2cm}|M{2cm}|M{2cm}|M{2cm}|| }
\hline
    \multicolumn{2}{||c|}{\textbf{Requirement}}& \textbf{Pre-VR} & \textbf{Entry-Level VR} & \textbf{Advanced VR} & \textbf{Human Perception} & \textbf{Ultimate VR}  \\ \hline
    \multicolumn{2}{||c|}{\textbf{Experience Duration}} & less than 20 minutes & less than 20 minutes & less than an hour & / & more than an hour  \\ \hline
    \multicolumn{2}{||c|}{\textbf{Video Resolution}} & 3840$\times$1920 (Full-view 4K Video) & 7680$\times$3840 (Full-view 8K Video) & 11520$\times$5760 (Full-view 12K Video) & 21600$\times$10800 (Full-view Video) & 23040$\times$11520 (Full view 24K Video) \\ \hline
    \multicolumn{2}{||c|}{\textbf{Single-eye Resolution}} & 1080$\times$1080 & 1920$\times$1920 & 3840$\times$3840 & 9000$\times$8100 & 9600$\times$9600 \\ \hline
    \multicolumn{2}{||c|}{\textbf{Field-of-View (Single-eye)}} & 100$\times$100 & 110$\times$110 & 120$\times$120 & 150$\times$135 & 150$\times$150 \\ \hline
    \multicolumn{2}{||c|}{\textbf{Bit per Color (RGB)}} & 8 & 8 & 10 & / & 12 \\ \hline
    \multicolumn{2}{||c|}{\textbf{Refresh Rate}} & 60 & 90 & 120 & 120 & 200 \\ \hline
    \multicolumn{2}{||c|}{\textbf{Pixel per Degree}} & 10 & 17 & 32 & 60 & 64 \\ \hline
    \multirow{5}{1em}{\\ \textbf{Service Require-ment}} & \textbf{Uncompressed Bit Rate (Progressive 1:1)*} & 10.62 Gbps & 63.70 Gbps & 238.89 Gbps & 1007.77 Gbps & 1911.03 Gbps \\ \cline{2-7} 
     & \textbf{Transmitting Bit Rate (Low-latency Compression 20:1)} & 530 Mbps & 3.18 Gbps (Full-view)  796 Mbps (FoV) & 11.94 Gbps    (Full-view)  5.31 Gbps (FoV) & 50.39 Gbps (Full-view)  31.49 Gbps (FoV) & 95.55 Gbps (Full-view)  66.36 Gbps (FoV) \\ \cline{2-7}
     & \textbf{Transmitting Bit Rate (Lossy Compression 300:1)} & 35 Mbps & 210 Mbps (Full-view)    53 Mbps (FoV) & 796 Mbps (Full-view)    354 Mbps (FoV) & 3.36 Gbps (Full-View)    2.10 Gbps & 6.37 Gbps (Full-view)    4.42 Gbps (FoV) \\ \cline{2-7}
     & \textbf{Typical Round Trip Time (RTT)} & $10$~ms & $10$~ms & $5$~ms & $10$~ms & $5$~ms \\ \cline{2-7}
     & \textbf{Typical Packet Loss} & $10^{-6}$ & $10^{-6}$ & $10^{-6}$ & $10^{-6}$ & $10^{-6}$ \\ \hline
     \multicolumn{7}{l}{* Progressive Data rate = (3$\times$Bit per Color) $\times$ (Pixel per Degree$\times$Field-of-view (Full-view or Single-eye)) $\times$ Refresh Rate $/$ Compression ratio}
\end{tabular}
\end{table*}

\subsubsection{VR Interaction Latency}
The  VR interaction latency is one of the key requirement in VR service and is defined as the time starting from the user's movement to the time where the virtual environment responds to the user's movements. For non-VR application, the interaction latency cannot exceed $100$~ms for action like pressing the key \cite{Cuervo2018}; for VR application, the VR interaction latency should be less than  $20$~ms to avoid motion sickness and discomfort. The VR interaction latency is also significantly impacted by the human brain and perception \cite{Saad2019}. To cater to different requirements of human and applications, it is common to define the minimum VR interaction latency as $10$~ms \cite{Bastug2017a}.

\subsection{Mapping Human Perception Requirements to QoS}
Given the limits of human perceptions, we can map them to concrete VR  QoS requirements, such as data rate, error rate, and communication delay.
\subsubsection{Data Rate}
To satisfy the human perception requirements with 60 PPD minimum resolution and  \SI{120}{\hertz} minimum refresh rate, the minimum downlink data rate will be around 50.39 Gbps for 360 degrees VR transmission under a 20:1 compression rate as shown in Table. \ref{requiremnt-different-phase}. Note that, with a higher compression rate, the data rate requirement can be reduced by sacrificing the VR interaction latency using more processing resources. The data rate of uplink motion information is around 100-150 kbps, which is negligible compared to the downlink video streaming \cite{Tan2018}.

\subsubsection{VR Interaction Latency}
The $10$~ms minimum VR interaction latency requirement is not restricted to the communication delay, but it also includes the rendering delay and the video coding/decoding latency. Meeting this unique VR delay requirement will, in turn, lead to a very stringent constraint on the wireless communication latency which must now be much smaller than $10$~ms. As an example, the VR test on a Huawei 5G network with cloud service \cite{Zhang} showed that the communication latency accounts for $17.9$~ms out of the $82.2$~ms VR interaction latency, as shown in  Fig. \ref{fig:systemStruc}. Even with prediction, only one-third of the target VR interaction latency is available for wireless transmission, which further strains the resources of the cellular network. Under such stringent latency constraints, few additional milliseconds in the communication latency can have a significant influence on the VR QoS.

%

\subsubsection{Error Rate}
VR devices support two major types of traffic: motion information and VR stream data. The motion information requires zero error. As the visual environment is constructed based on the real-time user's motion and location information, an error in motion information can require an additional processing delay in order to regenerate the error pixels \cite{Chen2018}. Thus, it is desirable to maintain the error rate of VR stream data below $10^{-6}$ to avoid lost,  degraded, or damaged frames detectable by a human. Even with a lower error rate, the VR frame error may propagate and be visible over many consecutive frames.

\subsection{Mapping VR Phases to Quality-of-Service}
Huawei has defined \emph{four
evolved VR phases} with different VR device technical specifications: Pre-VR, Entry-level VR, Advanced
VR, and Ultimate VR phase \cite{HuaweiiLab2017}.
 Each phase defines a new level of VR device technical specification, which is primarily related to the development of the VR device. Based on these specifications, we can calculate the service requirement at each phase as shown in Table \ref{requiremnt-different-phase}. Clearly, the current commercial VR devices in Table \ref{headset-parameter} fall in the pre-VR and Entry-level VR phases, and the service requirement governed by human perceptions lies between the Advanced VR and the Ultimate VR phases.

\begin{figure}[t]
    \centering
    \includegraphics[width = 0.4\textwidth]{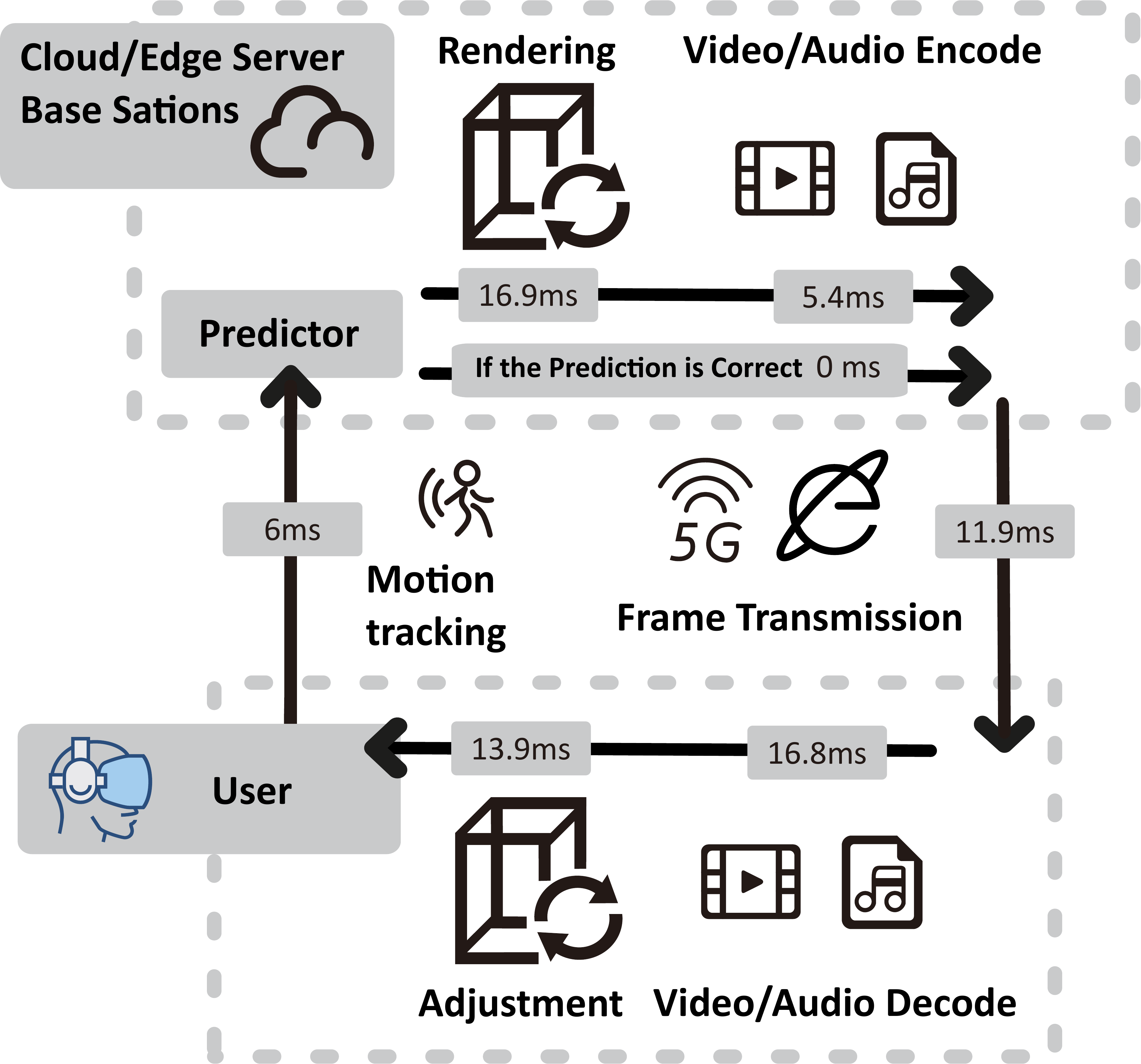}
    \caption{Decomposition of end-to-end VR interaction latency in Huawei 5G Test with cloud server \cite{Zhang}.}
    \label{fig:systemStruc}
\end{figure}

\begin{figure*}[ht]
    \centering
    \includegraphics[width = \textwidth]{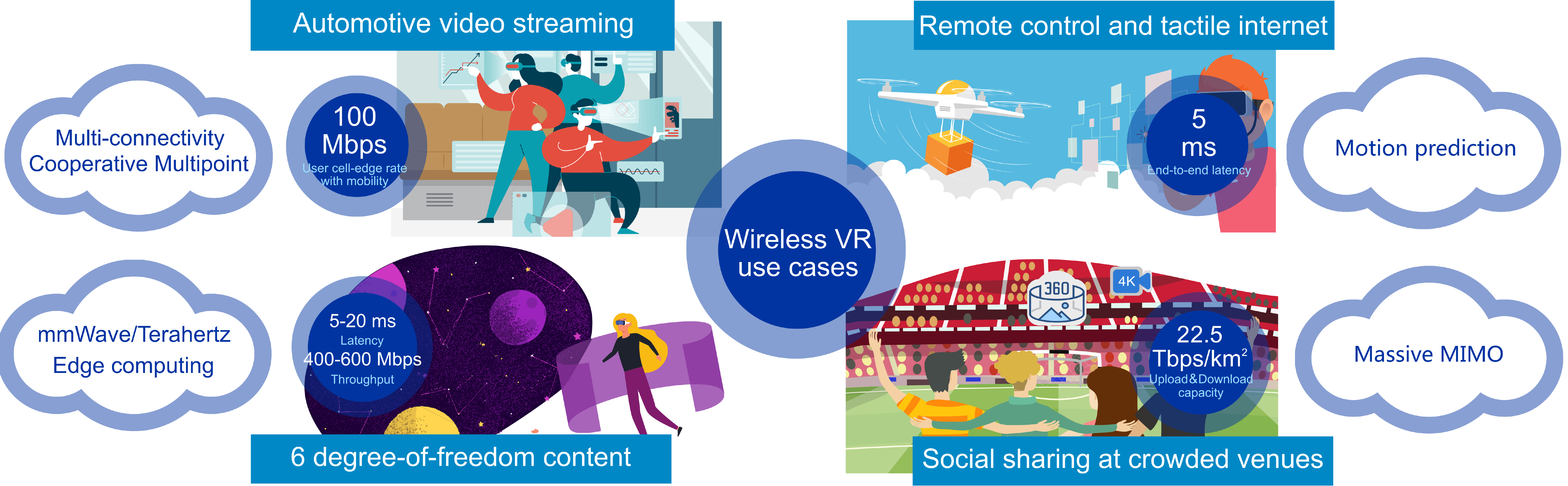}
    \caption{Cellular-connected wireless VR use cases, challenges, and potential solutions in 5G \cite{Quacom}. The data rate and latency values are suggested in the report which is based on the existing commercial VR devices in Entry-level VR phase as shown in Table \ref{requiremnt-different-phase}.}
    \label{fig:5guc}
\end{figure*}

\section{Use Cases, Challenges, and Potential Solutions}

Based on the VR requirements of Section II, we identify  several key characteristics of
VR traffic that makes it different from conventional video
traffic:
\begin{itemize}
    \item VR applications require a seamless low delay high capacity guarantee service for real-time rendering and VR interaction latency;
    \item VR content changes following the real-time users' motion;
    \item The uplink and downlink data rate requirements are asymmetric and coupled as they are both correlated to the end-to-end VR performance.
\end{itemize}

The aforementioned unique characteristics of VR services
bring forward new research challenges for existing wireless networks. For outdoor VR applications, LTE has been tested in \cite{Tan2018} under various mobility and signal strength conditions, and it is shown to support around 15 Mbps traffic with $10-30$~ms network delay, which does not even reach the pre-VR phase requirement in Table \ref{requiremnt-different-phase}. It is anticipated that 5G systems will deliver much higher data rates than LTE and, hence, it can better support ubiquitous connectivity for VR services. It is reported in \cite{NGMNAlliance2015} that the minimum downlink and uplink targets for 5G will be more than 50 Mbps everywhere, and the maximum downlink data rate at the user can be up to 1 Gbps, which only reaches the data rate requirements of the Advanced VR in Table \ref{requiremnt-different-phase}.

By quantifying the  capacity, latency, and reliability requirements for cellular-connected VR services, a recent technical report from \cite{Quacom} has classified  cellular-connected VR applications into four main use cases, which are automotive video streaming (VR-AVS), social sharing at crowded venues (VR-SS), 6 degree-of-freedom content streaming (VR-DoF), and remote control/tactile Internet (VR-RC), as shown in 
Fig. \ref{fig:5guc}. In the following, these four use cases with their service requirements are described followed by their key challenges and potential solutions.

\subsection{Automotive Video Streaming}

The automotive video streaming (VR-AVS) use case will allow commuters to enjoy live VR video streaming in high-speed trains and cars, which leads to technical challenges pertaining to supporting uniform and seamless connectivity with 
100 Mbps data rate (Early-level VR phase) for high-mobility VR users using cellular networks.
Next, we list potential solutions to cope with cellular network challenges brought by high mobility.   


\subsubsection{Dual Connectivity}
Dual connectivity, which allows multiple base stations (BSs) to transmit and receive data simultaneously to and from VR-AVS users, provides new opportunities for
maintaining seamless VR service. \textcolor{black}{Dual connectivity has been introduced as an important feature in LTE-Advanced (Rel-10/11) as well as part of 5G Multilink.}
For delay and resource sensitive VR applications, dual connectivity can reduce the service glitch caused by resource shortage, blockage, handover failure, \textcolor{black}{and poor cell-edge performance.}
\textcolor{black}{For instance, a VR-AVS video can be split between two BSs via drastically different bands for uplink and downlink
communications for more efficient and robust transmission via inter-cell resource aggregation. As a result, part of the video can be received if errors occur over one of the links.} Specifically, a sub-\SI{6}{\giga\hertz} band with decent coverage can be suitable for uplink
motion information (small packets), and the millimeter-wave (mmWave) band that provides  high data rates can be used for downlink
VR video traffic. \textcolor{black}{However, for users with high mobility in VR-AVS use case, the association, energy, and resource allocation are the major problems for dual connectivity. The backhaul network also requires to re-route the required traffic following the movement of VR users.}
\subsubsection{Coordinated Multipoint Transmission and Reception}
\textcolor{black}{Under the aforementioned conditions, coordinated multipoint transmission (CoMP), can be a suitable solution for addressing the challenge related to possible fluctuations in the data rates due to blockage (under mobility), by serving VR-AVS users using multiple BSs (multiple-path) over the same spectrum at the same time. Note that CoMP has been listed as a key feature in LTE-Advanced and 5G because of its ability to support delay-sensitive wireless applications \cite{3GPP2013}.} For VR applications, the downlink data rates can be improved to continuously meet the QoS needs of the VR users and the uplink motion information can be jointly processed for higher reliability to ensure accurate motion tracking.



\subsection{Remote Control and Tactile Internet}

The remote control and tactile internet (VR-RC) use case pertains to applications that involve a
remote control of cellular-connected unmanned-aerial-vehicles (UAVs) and robots over very long distances using VR. In this use case, the main challenge is how to support seamless connectivity with ultra-low end-to-end latency (e.g., within $5$~ms latency). \textcolor{black}{Similar to other delay sensitive applications, such as vehicular networks, the existing solutions for delay reduction include caching, network slicing and edge computing etc.}
In the latency-sensitive VR-RC use case, motion prediction
using machine learning can largely benefit the VR user experience at multiple levels. For instance, the prediction of users' motion allows pre-fetching the FoV of the users, as shown in Fig. \ref{fig:systemStruc}. However, it is important to note that the system needs to be able to tolerate prediction inaccuracy and to re-generate the correct content within the delay threshold \cite{Chen2018}. \textcolor{black}{It is also important to ensure the generalization of these prediction algorithms and enable transfer learning for other scenarios and applications.}
Second, motion prediction on the users' movement and location can be also beneficial for the quick association of VR users and remotely-controlled robots. Recall that VR services that use high-frequency bands are highly vulnerable to blockage due to obstacles. 
It is possible to proactively associate users by predicting the orientation and mobility patterns of VR users with known environment information.
In this way, sudden drops in the control link's QoS caused by blockage can be reduced.


\subsection{The 6 Degree-of-freedom Content Streaming}

The 6 degree-of-freedom (VR-DoF) content streaming use case will allow users to intuitively interact with a high quality immersive virtual world while moving freely inside the virtual environment. This use case requires meeting unique challenges related to supporting high data rates ($400-600$~Mbps, Advanced VR stage), and low latency  ($5-20$~ms) with accurate positioning of the VR user. These challenges can be addressed using the following solutions. 



\subsubsection{High Frequency Transmission}
In VR-DoF, the immersive experience of VR user must be supported through high-resolution video transmission, imposing stringent bandwidth requirements. The desire for higher video compression rate and lower compression latency calls for additional radio resources. As shown in Table \ref{requiremnt-different-phase}, the Ultimate VR phase with a $20$:$1$ compression rate under ultra-low latency constraints requires more than $100$ Gbps throughput. In this context, high-frequency transmission like mmWave ($30-300$ \SI{}{\giga\hertz}) and terahertz (\SI{}{T\hertz}) (\SI{100}{\giga\hertz}--\SI{10}{T\hertz}) communication are promising technologies to deliver VR-DoF content for wearable VR devices due to their high bandwidth availability and small form factor \cite{Chaccour2019}. \textcolor{black}{Using the aforementioned technologies, Vive has already implemented a wireless adapter for delivering such service via a mmWave connection from two access points within a small open room.} 

However, the main drawback of high-frequency communication is that it is unreliable and cannot maintain a high data rate over non-line-of-sight (NLoS) channels. This is because the penetration loss increases with increasing the frequency for most solid materials. In fact, even the human body can block the signal and cause glitches in the data stream. 
Overcoming this challenge requires new techniques, including the use of a reflector for re-establishing line-of-sight connectivity, proactive mobility management, dynamic frame structure, and integration of mmWave with sub-6 GHz frequencies \cite{Semiari2019,Chen2018}.

\subsubsection{Edge Computing}
VR-DoF content streaming VR applications will require effective computing to execute tasks such as rendering the VR environment, processing user motions, and computing data in the virtual environment. To maintain a low computing delay and avoid end-to-end transmission latency to a remote cloud, it is desirable to rely on edge computing techniques in which video rendering and computations are executed at edge devices that include BSs, edge servers, or even the VR devices themselves.
Determining where (cloud or edge) and how to perform computing tasks for VR purposes is therefore an important challenge. Performing computations at the level of the VR devices can significantly reduce the transmission latency. However, the limited computing capacity of end-user VR devices can lead to higher computing latency. As such, it is of interest to study how one can distribute computing tasks to different edge devices. Overcoming these challenges requires research across computing and communication domains. For instance, in \cite{Chen2018}, the author have developed several video transmission and caching schemes that take into account both the processing and transmission aspects of VR networking. In the commercial area the overall latency with state-of-the-art image processing and networking for VR was analyzed in \cite{HuaweiiLab2017} and by Google Stadia which promises to achieve $100$~ms end-to-end edge/cloud gaming experience.


\subsection{Social Sharing at Crowded Venues}

The social sharing at crowded venues (VR-SS) use case will allow a large group of users to share a VR experience in a public area, such as a bar or stadium. The main challenges facing the VR-SS use case include the need to support high uplink/downlink data rates with $12.5$ Tbps/km$^2$ in each direction and the need to integrate diverse VR applications with different service requirements \cite{Quacom}. It is a critical challenge for the cellular network to support a broad range of bandwidth-consuming and latency-critical VR services with various QoS requirements.




To guarantee a target perceptual resolution, massive MIMO can be employed to exploit 
spatial diversity and provide the multi-gigabit data rates
required by static and mobile VR-SS users over large coverage areas. 
The large antenna array of massive MIMO systems will also provide more angle-of-
arrival information, which  benefits the positioning
the system in VR. \textcolor{black}{Although many prior research has already shown that massive MIMO is effective to provide cellular connectivity, there is still a need for enhancing the stringent VR QoS requirement, especially for ultra-dense device connectivity in VR-SS use case. Beyond increasing the number of antennas within a limited space, it is also important to look for an economically viable approach to serve a large public crowd.}
For an indoor VR-SS environment with dense VR users in an office building, holographic
massive MIMO using a large intelligent surface can
potentially be used to overcome unfavourable propagation conditions
or to enrich the channel by introducing multi-paths,
through integrating an uncountable and infinite number of
antennas into a limited surface area.


\section{Case Study}
In light of the aforementioned VR challenges, it is necessary to identify the joint high reliability, 
low latency, as well as high data rate requirements
for all four cellular-connected VR use cases. 
More importantly, there is a fundamental difference between a traditional video service (even a 360$\degree$ video) and a VR-DoF streaming service due to the new human perception requirements in VR that we discussed in Section II.  While a streamed 3DoF video
(i.e., 360$\degree$ video) content can tolerate unstable QoS via a jitter buffer, the rendering of 6DoF
content calls for real-time transmission
with low interaction latency \cite{Quacom}.

To compare the rate-reliability-latency performance
between VR-DoF streaming and traditional video streaming in cellular networks,
we examine the percentage of successfully delivered bitplanes within a $7$~ms constraint between traditional video transmission and VR video transmission
in the same wireless mmWave network setting. \textcolor{black}{To separate the $10$~ms delay constraint into uplink and downlink, we have distributed the values of the delay constraint according to the proportion of downlink delay in the context of the round-trip delay ratio in Fig. \ref{fig:systemStruc}, which is $7$~ms.} Both types of video are transmitted following a MPEG video coding scheme, which separates a video stream into bitplanes, each of which being of size $1,578$~bits, calculated based on a $768$~Mbps bit rate requirements for the Advanced phase in Table \ref{requiremnt-different-phase}. Then, we apply dual connectivity (with two mmWave links) for VR-DoF streaming to show the need for new technologies to deliver such service, \textcolor{black}{where the second-closest base station acts as the fixed second base station, which transmits the same data simultaneously in another mmWave band.}
In this regard, we conduct a VR-DoF use case study with $N_\text{u}$ randomly distributed viewers
using the 2D map of King's College London within a
simulation environment with three BSs (with fixed locations at ME London, Strand Building, and Somerset House). \textcolor{black}{For the wireless connection, we use a classical 3GPP mmWave channel model \cite{3GPP2018a} operating at \SI{28}{\giga\hertz} taking into account the penetration loss and the indoor path loss through real building walls in King's College London. We also consider two types of schedulers at the BSs, namely proportional fair (PF) or round-robin (RR) scheduler following their implement in NS-3.}

\begin{figure}[tb]
    \centering
    \includegraphics[width = 0.5\textwidth]{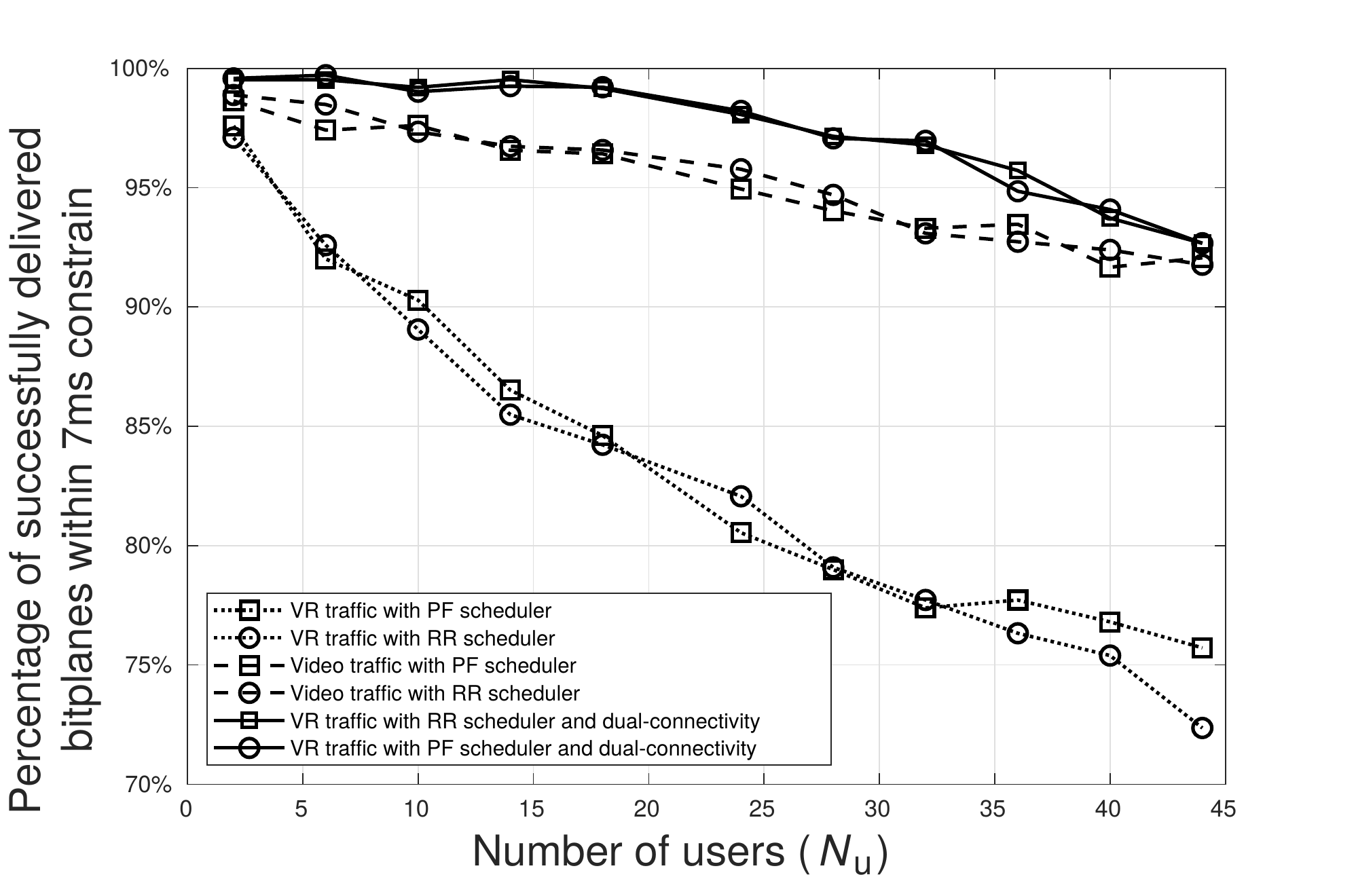}
    \caption{The percentage of successful delivered bitplanes with a delay constraint of $7$~ms for VR and traditional videos.}
    \label{fig:usernum}
\end{figure}


To measure the rate-reliability-latency tradeoff, in Fig. \ref{fig:usernum}, we plot the percentage of successfully delivered bitplanes within a $7$~ms downlink delay constraint versus the number of users for VR streaming and traditional video during within a transmission time of $5$ minutes.
As expected, this metric for both traffic types decreases with increasing the number of users to a larger cell load and higher interference. Surprisingly, the
performance of the two schedulers with the same type of traffic
are similar due to the relatively low downlink traffic load. We also notice that 
the cellular network failed to support VR streaming effectively in the rate-reliability-latency space, as can be seen from a much lower percentage of successfully delivered bitplanes within $7$~ms constraint compared with traditional video streaming. This is because the uncertainty of the channel conditions in mmWave networks result in performance degradation, that is unbearable for VR user experience.  Importantly, it is also shown that our proposed solution -- dual connectivity -- can largely combat the unstable wireless channel by connecting a VR device with two BSs simultaneously to provide enhanced performance.
 This case study clearly highlights the need for providing seamless VR connectivity within rate-reliability-latency space by using new approaches such as dual connectivity.

\section{Conclusion}
In this paper, we have envisioned a new paradigm for integrating VR into wireless cellular networks for providing seamless immersive VR applications from anywhere at any time. From a VR user perspective, we have first defined the QoS requirements for four wireless VR technological phases based on human perception requirements. We have then presented four use cases and corresponding potential solutions for cellular-connected VR communication by quantifying their specific QoS requirements and potential trade-off. Importantly, the case study has demonstrated that a cellular mmWave network can potentially support the VR streaming with new QoS requirement via the help of dual connectivity, and identified the unique QoS requirements of VR transmission compared with that of traditional video service. This work serves to inspire research to customize optimization solutions for cellular-connected VR use cases with technologies like motion prediction, massive MIMO, and edge computing.

\vskip -2\baselineskip plus -1fil

\begin{IEEEbiographynophoto}{Fenghe Hu}
is currently a PhD student in the center of telecommunication research, King's College London. 
\end{IEEEbiographynophoto}

\vskip -2\baselineskip plus -1fil

\begin{IEEEbiographynophoto}{Yansha Deng} 
is currently a  Lecturer  (Assistant  Professor)  with the Department of Informatics, King’s College London. Her research  interests include  molecular communication, Internet of Things, and 5G wireless networks. 
\end{IEEEbiographynophoto}

\vskip -2\baselineskip plus -1fil

\begin{IEEEbiographynophoto}{Walid Saad} 
is a Professor at the Department of Electrical and Computer Engineering at Virginia Tech. His research interests include wireless networks, machine learning, game theory, cybersecurity, unmanned aerial vehicles, and cyber-physical systems. 
\end{IEEEbiographynophoto}

\vskip -2\baselineskip plus -1fil

\begin{IEEEbiographynophoto}{Mehdi Bennis}
is an Associate Professor at the Centre for Wireless Communications, University of Oulu, Finland. His main research interests are in radio resource management, game theory and machine learning in 5G/6G networks. 
\end{IEEEbiographynophoto}

\vskip -2\baselineskip plus -1fil

\begin{IEEEbiographynophoto}{Hamid Aghvami}
joined the academic staff at King’s College London in 1984. In 1993 was promoted Professor in Telecommunications Engineering. He is the founder of the Centre for Telecommunications Research at King’s. 
\end{IEEEbiographynophoto}


\begin{thebibliography}{10}
\providecommand{\url}[1]{#1}
\csname url@samestyle\endcsname
\providecommand{\newblock}{\relax}
\providecommand{\bibinfo}[2]{#2}
\providecommand{\BIBentrySTDinterwordspacing}{\spaceskip=0pt\relax}
\providecommand{\BIBentryALTinterwordstretchfactor}{4}
\providecommand{\BIBentryALTinterwordspacing}{\spaceskip=\fontdimen2\font plus
\BIBentryALTinterwordstretchfactor\fontdimen3\font minus
  \fontdimen4\font\relax}
\providecommand{\BIBforeignlanguage}[2]{{%
\expandafter\ifx\csname l@#1\endcsname\relax
\typeout{** WARNING: IEEEtran.bst: No hyphenation pattern has been}%
\typeout{** loaded for the language `#1'. Using the pattern for}%
\typeout{** the default language instead.}%
\else
\language=\csname l@#1\endcsname
\fi
#2}}
\providecommand{\BIBdecl}{\relax}
\BIBdecl

\bibitem{Quacom}
\BIBentryALTinterwordspacing
Qualcomm, ``{VR} and {AR} pushing connectivity limits,'' Qualcomm Technologies.
  Inc., Tech. Rep., 2018 (Accessed on 2019-12-19). [Online]. Available:
  \url{https://www.qualcomm.com/invention/extended-reality/virtual-reality}
\BIBentrySTDinterwordspacing

\bibitem{Tan2018}
Z.~Tan, Y.~Li, Q.~Li, Z.~Zhang, Z.~Li, and S.~Lu, ``Supporting mobile {VR} in
  {LTE} networks: How close are we?'' \emph{in Proc. ACM Meas. Anal. Comput.
  Syst.}, vol.~2, no.~1, pp. 1--31, March 2018.

\bibitem{Cuervo2018}
E.~Cuervo, K.~Chintalapudi, and M.~Kotaru, ``Creating the perfect illusion :
  What will it take to create life-like virtual reality headsets?'' \emph{Proc.
  19th Int. Work. Mob. Comput. Syst. Appl.}, pp. 7--12, February 2018, {NY},
  USA.

\bibitem{HuaweiiLab2017}
\BIBentryALTinterwordspacing
Huawei-iLab, ``Cloud {VR},'' Tech. Rep., 2017 (Accessed on 2019-12-19).
  [Online]. Available:
  \url{https://www-file.huawei.com/-/media/corporate/pdf/ilab/cloud_vr_oriented_bearer_network_white_paper_en_v2.pdf}
\BIBentrySTDinterwordspacing

\bibitem{Bastug2017a}
E.~Bastug, M.~Bennis, M.~Medard, and M.~Debbah, ``Toward interconnected virtual
  reality: Opportunities, challenges, and enablers,'' \emph{IEEE Commun. Mag.},
  vol.~55, no.~6, pp. 110--117, August 2017.

\bibitem{Elbamby2018a}
M.~S. Elbamby, C.~Perfecto, M.~Bennis, and K.~Doppler, ``Toward low-latency and
  ultra-reliable virtual reality,'' \emph{IEEE Netw.}, vol.~32, no.~2, pp.
  78--84, January 2018.

\bibitem{Saad2019}
\BIBentryALTinterwordspacing
W.~Saad, M.~Bennis, and M.~Chen, ``{A Vision of 6G Wireless Systems:
  Applications, Trends, Technologies, and Open Research Problems},'' \emph{IEEE
  Netw.}, 2020. [Online]. Available: \url{http://arxiv.org/abs/1902.10265}
\BIBentrySTDinterwordspacing

\bibitem{Chen2018}
M.~Chen, W.~Saad, and C.~Yin, ``Virtual reality over wireless networks:
  Quality-of-service model and learning-based resource management,'' \emph{IEEE
  Trans. Commun.}, vol.~66, no.~11, pp. 5621--5635, June 2018.

\bibitem{3GPP201807}
3GPP, ``Virtual reality ({VR}) media services over {3GPP} ({3GPP} {TR} 26.918
  version 15.2.0 release 15),'' 3GPP, Tech. Rep., 2018.

\bibitem{Zhang}
\BIBentryALTinterwordspacing
W.~Zhang, ``Cloud {X}: New services in {5G} era,'' Huawei, Tech. Rep., 2018
  (Accessed on 2019-12-19). [Online]. Available:
  \url{https://www.gsma.com/futurenetworks/wp-content/uploads/2019/03/Huawei-5G-PDF.pdf}
\BIBentrySTDinterwordspacing

\bibitem{NGMNAlliance2015}
\BIBentryALTinterwordspacing
{R. El Hattachi and J. Erfanian}, ``{NGMN} 5{G} white paper,'' NGMN Alliance,
  Tech. Rep., Feb. 2015 (Accessed on 2019-12-19). [Online]. Available:
  \url{https://www.ngmn.org/wp-content/uploads/NGMN_5G_White_Paper_V1_0_01.pdf}
\BIBentrySTDinterwordspacing

\bibitem{3GPP2013}
3GPP, ``{3GPP} {TR} 36.819: Coordinated multi-point operation for {LTE}
  physical layer aspects(release 11),'' 3GPP, Tech. Rep., 2013.

\bibitem{Chaccour2019}
C.~Chaccour, R.~Amer, B.~Zhou, and W.~Saad, ``On the reliability of wireless
  virtual reality at {T}erahertz ({THz}) frequencies,'' in \emph{Proc. of 10th
  IFIP International Conference on New Technologies, Mobility and Security
  (NTMS)}, Canary Islands, Spain, June 2019.

\bibitem{Semiari2019}
O.~Semiari, W.~Saad, M.~Bennis, and M.~Debbah, ``{Integrated millimeter wave
  and sub-6 GHz wireless networks: A roadmap for joint mobile broadband and
  ultra-reliable low-latency communications},'' \emph{IEEE Wirel. Commun.},
  vol.~26, no.~2, pp. 109--115, February 2018.

\bibitem{3GPP2018a}
3GPP, ``Study on channel model for frequency spectrum above 6 {GHz} ({3GPP}
  {TR} 38.900 version 14.3.1 release 14),'' 3GPP, Tech. Rep., 2017.

\end{thebibliography}
\end{document}